\documentclass[twocolumn]{article}

\usepackage{graphicx} 
\usepackage{amsmath}
\usepackage{comment}
\usepackage{siunitx}
\usepackage{subfigure}
\usepackage{abstract}
\usepackage{url}

\begin{document}

\title{Computational Study of Density Fluctuation-Induced Shear Bands Formation in Bulk Metallic Glasses}
\author{Siya Zhu,Hagen Eckert,Stefano Curtarolo,Jan Schroers,Axel van de Walle}

\maketitle

\begin{onecolabstract}
Seemingly identical Bulk Metallic Glasses (BMG) often exhibit strikingly different mechanical properties despite having the same composition and fictive temperature. A postulated mechanism underlying these differences is the presence of ``defects''. Here we investigate this hypothesis through the study of the effect of density fluctuations on shear band formation under an applied stress. We find that the critical shear stress is strongly dependent on the magnitude and size of the fluctuations. This finding also elucidates why, historically, critical shear stresses obtained in simulations have differed so much from those found experimentally, as typical simulations setups might favor unrealistically uniform geometries.

\end{onecolabstract}

\section{Introduction}
For decades, bulk metallic glasses (BMGs) have garnered significant attention due to their remarkable mechanical properties\cite{chen1970mechanical,schuh2007mechanical,eckert2007mechanical,yavari2007mechanical}. When subjected to external stresses, unlike in crystalline metals, localized shear phenomena are anticipated to predominate in BMGs, resulting in the formation of shear bands --- instances of plastic instability characterized by the localized occurrence of extensive shear strains within a relatively narrow band during material deformation\cite{donovan1981structure,shimizu2007theory,greer2013shear}. Shear bands hold considerable significance and garner widespread attention as they play a crucial role in unraveling the mechanics of deformation in BMGs\cite{liu2005initiation,miracle2011shear}. However, the formation of shear bands is not thermodynamically governed --- the dynamic process and the inhomogeneity in space complicate attempts at quantitative and accurate description of shear bands with experiments.

Molecular dynamics simulation emerges as an ideal tool for replicating and analyzing the shear banding that occurs during deformation. Unfortunately, Molecular Dynamics (MD) simulation comes with inherent limitations. Despite its capability to operate on a scale of millions of atoms and time steps, its scope still remains relatively modest compared to real-world processes, preventing the perfect reproduction of shear banding process in BMGs. Some previous MD work successfully reproduce the shear bands in BMGs, however, they either require significantly higher strains ($\epsilon>0.2$)\cite{zhou2016strengthening} or lower temperatures ($T\sim$ \SI{50}{\kelvin})\cite{cao2009structural} to induce the shear bands, or creates voids within the BMG structure to facilitate their formation\cite{zhou2016atomistic,zhou2018strengthening,yuan2025shear}.

In light of the observation that volumetric dilation accompanies with shear-band formation\cite{shimizu2007theory,cao2009structural,li2006atomic,shi2005strain}, we purposely introduce low-density regions in BMGs to quantify their impact on shear band formation in MD simulations. The low-density regions exhibit a realistic nature, potentially arising from impurities such as vacancies or non-metallic atoms, or originating from density fluctuations within the material. In this work, we use the well-known Zr$_{50}$Cu$_{50}$ BMG for our analysis of shear bands. We also define the critical strain for shear band initiation and analyze how the low-density areas affect the formation of shear bands.

\section{Results}

In this work, we randomly remove some atoms uniformly in a thin layer within a specified distance of a given $\left(\overline{1}01\right)$ plane (we use the concept of $\left(\overline{1}01\right)$ plane from crystalline structure to indicate the plane -x+z=0) of the structure, thus creating a low-density region along what will be the shear band direction and imitating plausible density fluctuations in a real BMG.
In Figure 1(b) and 1(c), we show the local shear of atoms under uniaxial strain of 0.1, while Fig 2 shows the average density in the structure versus the distance to the $\left(\overline{1}01\right)$ plane. 
With a low-density region along the $\left(\overline{1}01\right)$ plane, it is evident that the formation of shear bands is considerably facilitated.
\par

\begin{figure}[htp]
\centering
\includegraphics[width = \columnwidth]{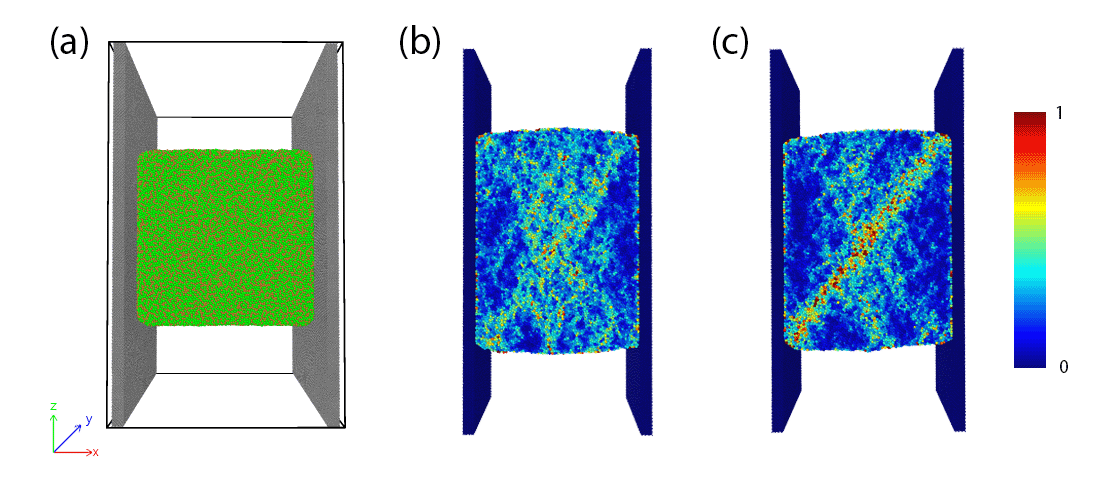}

\caption{\textbf{Atomic structures and local shear of BMG.}(a) Atomic structure with two diamond layers added on both sides of the ZrCu BMG to apply stress; (b) the local shear of atoms under strain $\epsilon_x = 10\%$ with perfect BMG Zr$_{50}$Cu$_{50}$; (c) the local shear of atoms under strain $\epsilon_x = 10\%$ with 3\% of the atoms within \SI{5}{\nano\meter} of the $\left(\overline{1}01\right)$ plane removed from the original structure to help forming the shear band.}  

\end{figure}

\begin{figure}[htp]
\centering

\includegraphics[width = \columnwidth]{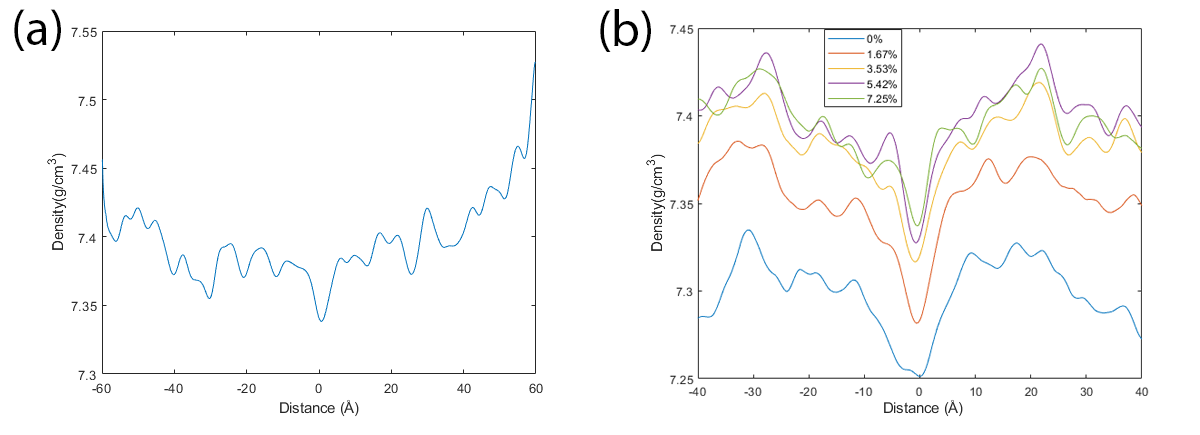}

\caption{\textbf{The average density of the Zr$_{50}$Cu$_{50}$ BMG as a function of distance to the $\left(\overline{1}01\right)$ plane.} (a) perfect BMG Zr$_{50}$Cu$_{50}$ under strain of 20\%; (b) 3\% of the atoms within \SI{5}{\nano\meter} of the $\left(\overline{1}01\right)$ plane are removed from the original structure under different strain.}

\end{figure}

Some prior studies of shear bands employed the concept of local shear by emphasizing the shear bands by marking atoms with $\mu_\text{vM}>0.2$ \cite{albe2013enhancing} or $0.28$\cite{feng2015atomic}. In this work, we employ a refined and precise methodology to delineate the shear bands by utilizing local shear strain data. 

\begin{figure}[htp]
\centering

\includegraphics[width = \columnwidth]{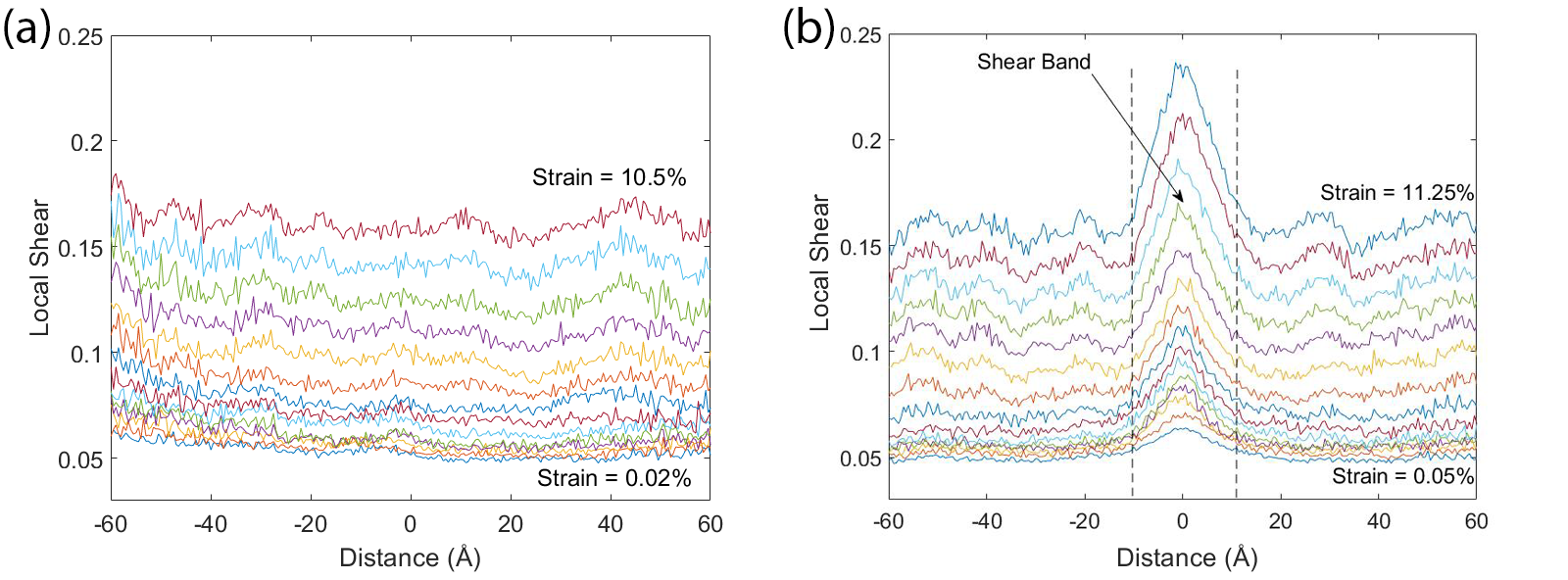}

\caption{\textbf{The average of local shear strains of atoms as a function of distance to the $\left(\overline{1}01\right)$ plane.} (a) perfect bulk  metallic glass Zr$_{50}$Cu$_{50}$. Strain from 0.02\% to 10.5\% is added on the X direction.; (b) 2\% of the atoms within \SI{5}{\nano\meter} of the $\left(\overline{1}01\right)$ plane are removed from the original structure to help forming the shear band. Strain from 0.05\% to 11.25\% is added on the X direction.}  

\end{figure}

In Figure 3, we plot the average of local shear strains of atoms as a function of the distance to the $\left(\overline{1}01\right)$ plane. We can see that the local shear strain as a function of the distance to the $\left(\overline{1}01\right)$ plane (or the shear band center) exhibits shifted Gaussian-like profile:
\begin{equation}
    \mu_{\text{vM}} = A e^{-\frac{x^2}{2c^2}}+b
\end{equation}
where $A$ is the height of the peak, $c$ is the standard deviation and $b$ is the baseline. We remove 0.5\% to 3\% of the atoms uniformly within the area of \SI{1}{\nano\meter}, \SI{3}{\nano\meter} and \SI{5}{\nano\meter} of the $\left(\overline{1}01\right)$ plane, applying strain along the X direction and plotting the A, c and b to describe the shear band, as shown in Figure 4. 
\begin{figure}[htp]
\centering
\includegraphics[width=\columnwidth]{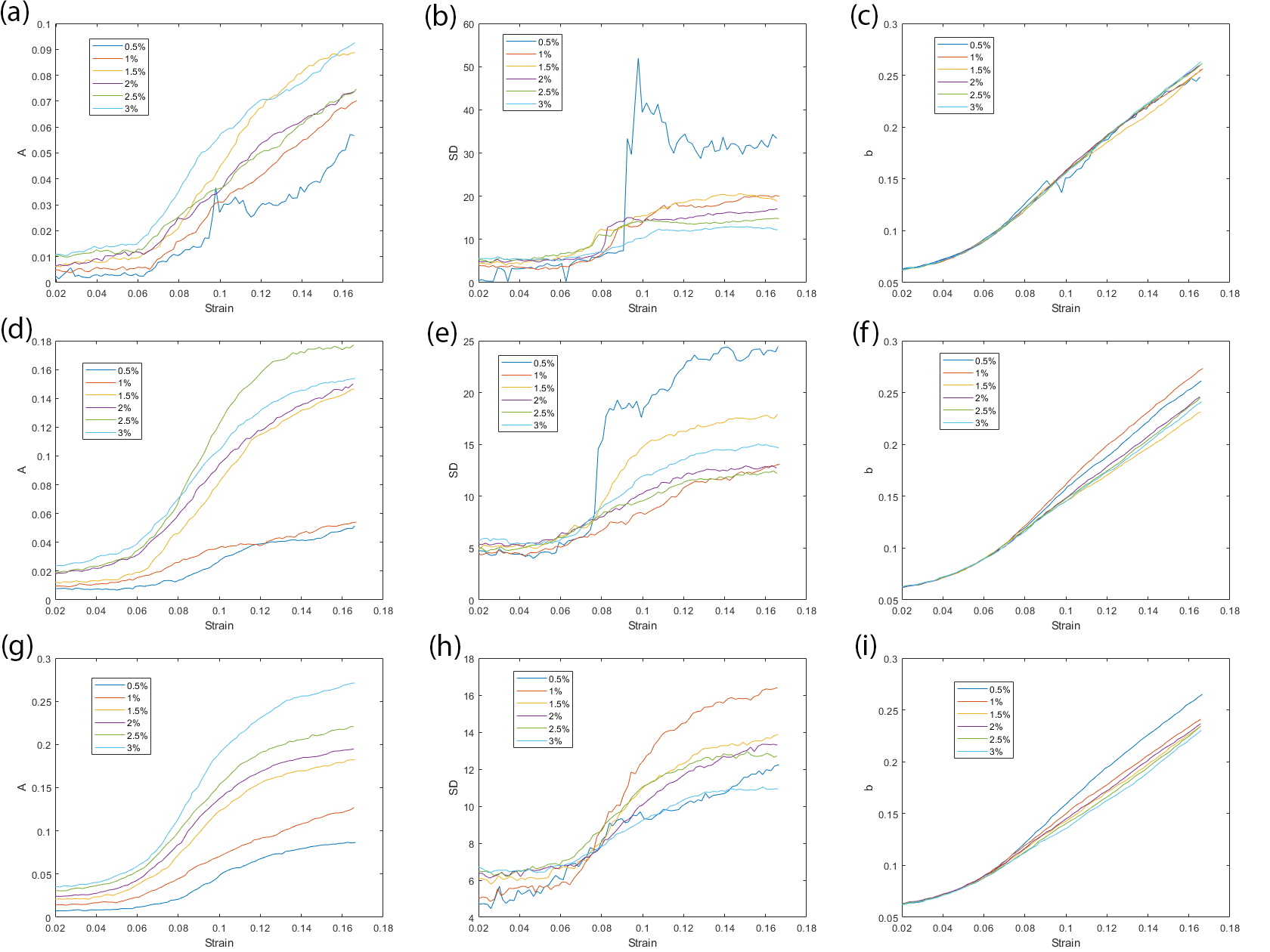}
\caption{\textbf{Parameters fitted as function of strain.} A, standard deviation and b fitted as a function of strain with different percentage of atoms removed within (a)-(c): \SI{1}{\nano\meter}; (d)-(f): \SI{3}{\nano\meter}; (g)-(i): \SI{5}{\nano\meter} of the $\left(\overline{1}01\right)$ plane.}  
\end{figure}
It is observed that A, the height of the Gaussian peak consistently exhibits an ``S-shaped'' relationship with respect to strain. (except in some cases with only 0.5\% atoms removed where no clear shear bands form) In other words, the second derivative of A to strain monotonically decreases from positive to negative in the strain range 0 to 0.18. We define the critical strain $\epsilon_\text{SB}$ for the shear band as the zero of the second derivative:
\begin{equation}
    \left(\frac{\partial^2{A}}{\partial \epsilon ^2}\right) \bigg| _{\epsilon = \epsilon_\text{SB}} = 0
\end{equation}
In Figure 5, we show the relationship between the critical strain and the percentage of atoms removed in the region. With an increase in the number of atoms removed from the shear band region, the critical strain for the shear band decreases to 0.07, which is much closer to experimental observations\cite{klaumunzer2010temperature}.
\begin{figure}[htp]
\centering
\subfigure{
\includegraphics[width = 0.8\columnwidth]{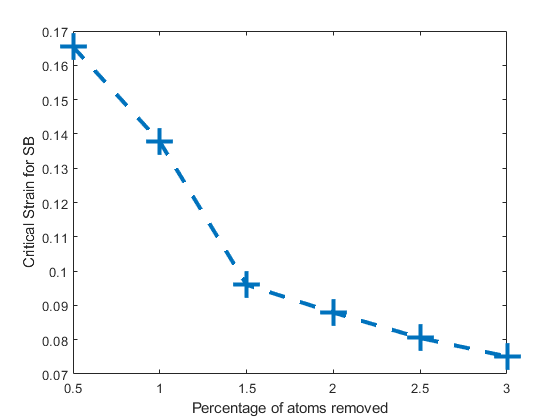}}
\caption{\textbf{Critical strain for shear band.} The critical shear banding strain for BMG with 0.5\% to 3\% atoms removed within \SI{3}{\nano\meter} from the $\left(\overline{1}01\right)$ plane are shown.}  
\label{fig8}
\end{figure}

\section{Discussion}
In this work, we propose a hypothesis regarding the influence of low-density regions on the formation of shear bands in bulk metallic glasses. We observe the presence of low-density regions within the shear band and successfully simulate shear band formation using EAM-MD at relatively low strain by removing no more than 3\% of atoms from a specific planar region of \SI{10}{\nano\meter} thickness. Our simulation setup does not require non-physical parameters, such as large uniaxial strain, low temperature, or the introduction of large voids in the structure. \par

This method enables a computational investigation of the physical behavior of shear bands in BMGs under more realistic conditions.
The shear banding process is non-steady in time and non-uniform in space. With the shear band atomic structures we generate, we develop an accurate and quantitative definition of shear bands and identify the critical strain for their formation in BMGs by utilizing a local shear parameter. Through the calculation of critical strain based on our MD simulations, we confirm the occurrence of shear banding at low strain.\par

In future work, our method can be extended to generate more complex multi-shear band structures, such as systems with groups of parallel shear bands, to further investigate their influence on the mechanical properties of BMGs. Additionally, we can apply our approach to systems with multiple elements and impurities, such as Zr-Cu-Al BMGs containing carbon or oxygen impurities (if reliable interatomic potentials are available), to explore how these impurities affect shear banding behavior in BMGs.

\section{Methods}
For all MD calculations, we use LAMMPS\cite{plimpton1995fast,thompson2022lammps} in conjunction with the EAM potentials from the group of Sheng\cite{cheng2009atomic}. For the unstrained BMG structure, we first perform a NVT and a NPT run at \SI{2000}{\kelvin}, and quenching to \SI{300}{\kelvin} at $10^{12}$ K/s with a NPT ensemble. Then a NPT at 300 K is performed to obtain the final stable structure. For the strained structures, two layers of diamond carbon are added on both sides, as shown in Figure 1(a). The interactions between carbon atoms and the BMG are simulated with Lennard-Jones potentials\cite{jones1924determination}. The carbon layers are forced to move into the middle along the X-axis to apply uniaxial strain upon the glass. This geometry is better suited to study shear band formation than an homogenous simulation with periodic boundary conditions, because it induces a non-uniform stress, reflecting the fact, in a real material, shear bands form in regions of non-homogenous stress. \par

The manifestation of shear bands within the metallic glass is elucidated through the assessment of the von Mises parameter, or the local shear strain experienced by each atom\cite{shimizu2007theory}.  For each atom $i$ in the system, a transformation matrix $\boldsymbol{J}_i$ is determined by minimizing the mapping error between the current ($\boldsymbol{d}_{ji}$) and reference ($\boldsymbol{d}_{ji}^0$) configurations: 
\begin{equation}
    \sum_{j\in N_i^0} \lvert \boldsymbol{d}_{ji}^0 \boldsymbol{J}_i - \boldsymbol{d}_{ji} \rvert ^2 ,
\end{equation}
in which the configuration are described by atomic coordinates and where the reference is the system's configuration before strain is applied.
Then the local Lagrangian strain matrix can be computed as
\begin{equation}
    \boldsymbol{\eta}_i = \frac{1}{2}\left(\boldsymbol{J}_i \boldsymbol{J}_i^\text{T} - \boldsymbol{I}\right)
\end{equation}
and the local shear (or the von Mises parameter) is defined as

\begin{align}
    \mu_i^\text{vM} & = \sqrt{\eta_{yz}^2 +\eta_{xz}^2 +\eta_{xy}^2} \notag \\
    & +\frac{\left(\eta_{yy}-\eta_{zz}\right)^2 + \left(\eta_{xx}-\eta_{zz}\right)^2 +\left(\eta_{xx}-\eta_{yy}\right)^2}{6}
\end{align}
The local shear calculations and atomic structure visualizations are performed using Open Visualization Tool (OVITO).\cite{stukowski2009visualization} Fitting for parameters in Equation(1) are performed using MATLAB.\cite{statisticsandmachinelearningtoolboxdocumentation}

\section{Data Availability}
The datasets generated and analyzed during the current study are available in the BMG-ShearBand repository, \url{https://drive.google.com/drive/folders/1IysoorT1AlDjFdjcaPtzFhDBPQ9Rw7ye?usp=drive_link}.

\bibliographystyle{unsrt}

\section{Acknowledgements}
This study was funded by Office of Naval Research grants N00014-20-1-2225 and N00014-20-1-2200. Computational resources were provided by (i) the Center for Computation and Visualization at Brown University, (ii) the Extreme Science and Engineering Discovery Environment (XSEDE) through allocation TGDMR050013N, which is supported by National Science Foundation Grant No. ACI-1548562 and (iii) the Advanced Cyberinfrastructure Coordination Ecosystem: Services \& Support (ACCESS) program through allocation DMR010001, which is supported by National Science Foundation grants 2138259, 2138286, 2138307, 2137603, and 2138296. The funders played no role in study design, data collection, analysis and interpretation of data, or the writing of this manuscript. 

\section{Competing Interests}
All authors declare no financial or non-financial competing interests.

\end{document}